\documentclass{birkjour}

\usepackage{showlabels}
\usepackage{hyperref}

\usepackage{amsmath}
\usepackage{amssymb}
\usepackage{latexsym}
\usepackage{graphicx}
\usepackage{color}

\usepackage{fullpage}

\newcommand{\seclab}[1]{\label{sec:#1}}

\newcommand{\secref}[1]{Section~\ref{sec:#1}}

\newcommand{\FieldF}{\mathbb{F}}
\newcommand{\Natural}{\mathbb{N}}

\newcommand{\Depth}{\textrm{depth}}
\newcommand{\Size}{\textrm{size}}
\newcommand{\Degree}{\textrm{degree}}
\newcommand{\SDet}{\textrm{SymDet}}
\newcommand{\Det}{\textrm{Det}}
\newcommand{\Perm}{\textrm{Perm}}
\newcommand{\HC}{\textrm{HC}}

\newcommand{\Sign}{\textrm{sign}}
\newcommand{\Var}{\textrm{var}}
\newcommand{\poly}{\textrm{poly}}
\newcommand{\BoolP}{\textrm{BoolPart}}
\newcommand{\SM}{\textrm{SM}}

\newcommand{\VF}{\textrm{VF}}
\newcommand{\VP}{\textrm{VP}}
\newcommand{\VQF}{\textrm{VQF}}
\newcommand{\VQP}{\textrm{VQP}}
\newcommand{\VNP}{\textrm{VNP}}
\newcommand{\VNF}{\textrm{VNF}}
\newcommand{\VNC}{\textrm{VNC}}
\newcommand{\VSAC}{\textrm{VSAC}}
\newcommand{\VBP}{\textrm{VBP}}
\newcommand{\VBWBP}{\textrm{VBWBP}}
\newcommand{\BWBP}{\textrm{BWBP}}
\newcommand{\Ptime}{\textrm{P}}
\newcommand{\PSPACE}{\textrm{PSPACE}}
\newcommand{\NP}{\textrm{NP}}
\newcommand{\NC}{\textrm{NC}}
\newcommand{\SAC}{\textrm{SAC}}

\begin{document}

\title{Algebraic Complexity Classes}
\author{Meena Mahajan}
\address{The Institute of Mathematical Sciences, 
    CIT Campus, Chennai 600 113, India.}
\email{meena@imsc.res.in}
\date{Oct 15, 2013}

\begin{abstract} 
This survey describes, at an introductory level, the algebraic
complexity framework originally proposed by Leslie Valiant in 1979,
and some of the insights that have been obtained more recently.
\end{abstract}
\thanks{The idea for writing this survey came while the author was
  working on the Indo-French CEFIPRA-supported project 4702-1.}
\subjclass{Primary 68Q15; Secondary 68Q05} \keywords{algebraic
  complexity, circuits, formulas, branching programs, determinant,
  permanent}

\maketitle

\section{Introduction}
\seclab{intro} 

In this survey, I am going to try and describe the algebraic
complexity framework originally proposed by Leslie Valiant
\cite{valiant-stoc79a,valiant82}, and the insights that have been
obtained more recently. This entire article has an ``as it appeals to
me'' flavour, but I hope this flavour will also be interesting to many
readers. The article is not particularly in-depth, but it is an
invitation to read \cite{bcs-book97,burgisser-book00} and many recent
papers on the topic, and to start attacking the open problems in the
area.

Valiant started out with the mission of understanding the core essence
of reductions and completeness, as witnessed in both recursive
function theory and in computational complexity theory. He provided an
algebraic framework in which to interpet the clustering of natural
problems into completeness classes, even for problems of an algebraic
rather than combinatorial nature. He had a remarkable hypothesis:
\begin{quote}
Linear algebra offers essentially the only fast technique for
computing multivariate polynomials of moderate degree.
\end{quote}
Clearly, then, we are going to talk about polynomials, not languages
or functions. 

\section{Valiant's original framework}
\seclab{orig} 

Let $\FieldF$ be any field, and let $\FieldF[x_1, \ldots, x_n]$ be the
ring of polynomials over indeterminates $x_1, \ldots , x_n$ with
coefficients from $\FieldF$. Consider a family $(f)$ of polynomials
$(f_n)_{n \ge 1}$, where each $f_n$ is in $\FieldF[x_1, \ldots,
  x_{s(n)}]$ for some function $s:\Natural \longrightarrow
\Natural$. When should we say that $(f)$ is tractable?  Clearly, if
there are too many variables to keep track of, there cannot be
tractability. So we will henceforth demand that $s$ is a polynomially
bounded function ($\exists c, \forall n, s(n) \le c+n^c$); then the
$n$th polynomial $f_n$ has $O(n^c)$ variables. But that is of
course not enough.

There are many ways in which we can set the bar for
tractability. Here's a first attempt. Can $(f)$ be computed by a {\em
  formula} of reasonable \Size? To elaborate further, a formula is an
expression defined recursively: 
\begin{enumerate}
\item
for each $c\in \FieldF$, ``$c$'' is a formula of size 0 computing the
polynomial $c$,
\item for each indeterminate $x_i$, ``$x_i$'' is a formula of size 0
  computing the polynomial $x_i$, and
\item if $F_1, F_2$ are formulas computing polynomials $f_1$ and
  $f_2$, then ``$(F_1 + F_2)$'' and ``$(F_1 \times F_2)$'' are
  formulas of size $\Size(F_1) + \Size(F_2) + 1$ each, computing the
  polynomials $f_1+f_2$ and $f_1 \times f_2$ respectively.
\end{enumerate}
Notice that $\Size(F)$ is just the number of ring/field operations used to
construct $F$.  

Instead of such a recursive definition, we could have a more intuitive
picture: a formula is a rooted binary tree where internal nodes are
labeled $+$ or $\times$ and leaf nodes are labeled from the set
$\FieldF \cup X$, where $X$ is the set of indeterminates. The size is
just the number of non-leaf nodes. 

Now, for tractability, we could require that there is a polynomially
bounded function $t:\Natural \longrightarrow \Natural$ and a family of
formulas $(F_n)_{n \ge 1}$ such that each $F_n$ computes $f_n$ and has
size at most $t(n)$. Let us use the notation $\VF$ to denote families
of polynomials tractable in this sense. ($\VF$: Valiant's Formulas ---
of course, Valiant didn't use this name! This class is also referred
to as $\VP_e$: Valiant's Polynomial-sized Expressions. Personally, I
prefer \VF. Also note, in formal logic, the formulas/expressions
referred to above are called terms, hence $\VF$ means families with
polynomial ``termic complexity''. )

Here's a second attempt: Can $(f)$ be computed by a {\em straight-line
  program} of reasonable size? As before, we will declare polynomial
size to be reasonable. Straight-line programs are programs where
instructions involve adding or multiplying previously computed
polynomials, no divisions and no conditionals (no if-then-else). In
the more intuitive picture, they correspond to directed {\em acyclic}
graphs where each node is a source node (indegree 0) labeled from the
set $\FieldF \cup X$, or has indegree 2 and is labeled $+$ or
$\times$. A designated sink node (outdegree 0) is the output
node. Each node computes a polynomial in the obvious way, and the
graph computes the polynomial at the output node. (The acyclicity
constraint ensures that there is a linear ordering of the nodes such
that each node, or instruction, only uses
previously computed polynomials. This gives the straight-line
program.) The size is the number of non-source
nodes; again, this corresponds to the number of ring/field operations
required. Such graphs are in fact exactly {\em algebraic circuits},
and we now look for polynomial size circuit families.

Clearly, this model generalises formulas. The catch is that it
generalises it too much! To see why, consider the following circuit
family: $C_n$ has $n+1$ nodes $v_0, v_1, \ldots , v_n$, and the
labeling is $v_0=x_1$, $v_{i+1} = v_i \times v_i$ for $i\in [n]$. The
family of polynomials $(f_n)$ computed by $(C_n)$ is $f_n =
x_1^{2^n}$.  Even for small integer values of $x_1$, writing down the
value of $f_n(x_1)$ is going to require exponentially many bits. How
can we say that such a family $(f_n)$ is tractable?

So we need to impose some additional restrictions. The obvious
parameter to restrict is the degree of the polynomial. Say that the
family $(f_n)$ has moderate degree if for some polynomially bounded
function $d : \Natural \longrightarrow \Natural$, the \Degree\ of each
polynomial $f_n$ is at most $d(n)$. If $\Degree(f_n)=D$ is
polynomially bounded, then on integer arguments with $b$-bit
representations, the value of $f_n$ requires no more than $\poly(n,b)$
bits. (In general, it needs no more than $\poly(n,D,b)$ bits.)
 Henceforth, to qualify for the label tractable, a family
$(f_n)$ must have polynomially bounded degree.

(Why didn't we face this problem when considering $VF$? Simply because
a formula of \Size\ $t$ cannot compute a polynomial of degree more
than $t+1$. Don't just believe me; check this by induction on formula
\Size.)

Now we have our second possible definition of tractability: $(f_n)$ is
tractable if the sequence $\Degree(f_n)$ is polynomially bounded, and
there is a polynomially bounded function $t:\Natural \longrightarrow
\Natural$ and a family of straight-line programs, or algebraic
circuits, $(C_n)_{n \ge 1}$, such that each $C_n$ computes $f_n$ and
has size at most $t(n)$. Let us use the notation $\VP$ to denote
families of polynomials tractable in this sense. ($\VP$: Valiant's
analogue of the Boolean complexity class \Ptime. Valiant called these
families $p$-computable \cite{valiant82}.)

The well-studied polynomial family from linear algebra, the
determinant of a matrix of indeterminates, is known to be tractable in
this second sense.  (To define the family $(\Det_n)$, imagine an $n
\times n$ matrix $A_n$ with a new indeterminate $x_{ij}$ at each
position $(i,j)$, and let $\Det_n$ be the polynomial that represents
the determinant of $A_n$. Thus $\Det_1=x_{11}$,
$\Det_2=x_{11}x_{22}-x_{12}x_{21}$, and so on. Clearly, this family
satisfies the mandatory conditions: $\Det_n$ has $n^2$ variables and
is of degree $n$.)  This is not surprising; we know that the
determinant can be computed efficiently (in polynomial time) over
instantiated matrices using, say, Gaussian elimination. But to compute
the symbolic determinant via a straight-line program, Gaussian
elimination is apparently not directly of use because we can't search
for non-zero pivots and eliminate them (remember, no divisions and no
conditionals). However, Strassen \cite{strassen} gave a generic method
of converting any straight-line program with divisions to a
division-free straight-line program; the resulting program's size is
polynomially bounded in the original size, the number of variables,
and the degree. Thus we can conclude that there are polynomial-sized
straight-line programs for the symbolic determinant. There are more
direct algorithms as well; Samuelson, Berkowitz, Csanky, ... .  See
\cite{mahvin} for an explicit description of circuits of
size $O(n^4)$ (my favourite one -- no surprise!).

Whether the determinant can be computed efficiently by formulas (is
$\Det_n$ in \VF?) is still famously open. We know that it needs
formula size at least $\Omega(n^3)$, see \cite{Kalorkoti85}.  But we
do know that it can be computed by formulas of sub-exponential size
$2^{O(\log^2 n)}$. This can be shown in many different ways, one of
which we will look at a bit later, but the earliest demonstration of
this follows from Csanky's algorithm \cite{csa}, which uses binary
arithmetic operations and $O(\log^2 n)$ parallel time.  Thus if we use
quasi-polynomial ($2^{\log^c n}$ for some constant $c$) formula-size
as the defining property for tractability (giving a class that we can
call \VQF), then again the family $(\Det_n)$ has long been known to be
tractable. We could also use quasi-polynomial circuit size as the
defining property for tractability, giving a class that we can call
\VQP. But \VQP\ obviously contains \VP\ and \VQF, so $(\Det_n)$ is in
\VQP; nothing new there. (Note: in defining \VQF\ and \VQP, the
quasi-polynomial limit on formula or circuit size is over and above
the requirement that the degree and number of variables are
polynomially bounded.)

Does $\VP$ include essentially all interesting and natural polynomial
families? We do not know. In fact, there is a large list of such
polynomial families not known to be in $\VP$. The most natural one is
the permanent family $(\Perm_n)$ where $\Perm_n$ is the polynomial
representing the permanent of an $n \times n$ matrix $A_n$ of
indeterminates. It is tantalisingly similar to the determinant; just
the sign term is missing.
\[
\Det_n  =  \sum_{\sigma \in S_n} \Sign(\sigma) \prod_{i=1}^n x_{i\sigma(i)}  
\mbox{\hspace*{1in}}
\Perm_n  =  \sum_{\sigma \in S_n} \prod_{i=1}^n x_{i\sigma(i)} 
\]
Yet, it does not seem to be tractable. How ``intractable'' is it? 

Mirroring the definitions of the Boolean complexity classes
\Ptime\ and \NP, Valiant proposed a notion of $p$-definability in
\cite{valiant-stoc79a}. A polynomial family $(f_n)$ is $p$-definable
if it can be written as an exponential sum, over partial Boolean
instantiations, of another tractable family. Formally, a family
$(f_n)$ over $s(n)$ variables and of degree $d(n)$ is $p$-definable if
$s(n)$ and $d(n)$ are polynomially-bounded, as always, and further,
there exist a polynomially-bounded function $m$, and a family of
polynomials $(g_n)$ in $\VF$, such that $g_n$ has $s(n)+m(n)$
variables denoted $\{x_1, \ldots , x_{s(n)}, y_1, \ldots ,
y_{m(n)}\}$, and \[f_n(\tilde{x}) = \sum_{y_1=0}^1 \sum_{y_2=0}^1
\ldots \sum_{y_{m(n)}=0}^1 g_n(\tilde{x}, \tilde{y}). \] This looks
like an algebraic analogue $\sum \cdot \VF$ of the boolean class
$\exists \cdot F$, where $F$ is the class of languages decided by
polynomial-size formulas. But it is well-known that $\exists \cdot F =
\NP$, so this should be algebraic \NP. Later, Valiant redefined
$p$-definability (no, that is not a circular definition!) as
exponential sums of families in \VP, rather than \VF; that is, $\VNP =
\sum \cdot \VP$. For clarity, let us agree to temporarily refer to
these two definitions as \VNF\ (or $\VNP_e$) and \VNP. However, Valiant
\cite{valiant82} showed that these two classes are in fact the same,
so just \VNP\ will do. The proof involves showing that \VP\ is
contained in $\sum \cdot \VF$. And it is of course easier to show
upper bounds with the definition of \VNP\ rather than \VNF.

Now Valiant observed that not only $(\Det_n)$, even $(\Perm_n)$ is
$p$-definable. This should be similar to showing that the 0-1
permanent is in \#\Ptime, right? Almost. We are dealing with symbolic
polynomials, so we do not have the liberty of looking at an input value
and deciding what to do next. Still, the basic idea is the same. For a
statement $S$, let $[S]$ denote the 0-1 valued Boolean predicate that
takes value 1 exactly when $S$ is true. Then 
\begin{eqnarray*}
\Perm_n = \sum_{\sigma \in S_n} \prod_{i=1}^n x_{i\sigma(i)} 
&=& \sum_{Y \in \{0,1\}^{n\times n}} 
\left[\parbox{20mm}{\textrm{$Y$ is a 0-1 permutation    matrix}}\right] 
\cdot \prod_{i=1}^n \left(\sum_{j=1}^nY_{ij}x_{ij}\right) \\
~\left[\parbox{20mm}{\textrm{$Y$ is a 0-1 permutation    matrix}}\right] &=&
[\textrm{$Y$ has at least one 1 in each row}]\times \\
&&[\textrm{$Y$ has at most one 1 in each line}\\
&& ~~\textrm{(line =row or column)}]\\
&= & \left(\prod_{i=1}^n \sum_{j=1}^n Y_{ij} \right) 
\left(\prod_{
\begin{array}{c}
\scriptstyle{(i,j) \neq (k.m)}; \\[-2mm]
\scriptstyle{i=k ~\textrm{~or~}~ j=m}
\end{array}
} (1- Y_{ij}Y_{km}) \right) 
\end{eqnarray*}
Clearly, the polynomial family 
\[ 
g_n = \left(\prod_{i=1}^n \sum_{j=1}^n Y_{ij} \right) \left(\prod_{
\begin{array}{c}
\scriptstyle{(i,j) \neq (k.m)}; \\[-2mm]
\scriptstyle{i=k ~\textrm{~or~}~ j=m}
\end{array}
} (1- Y_{ij}Y_{km}) \right) \prod_{i=1}^n
\left(\sum_{j=1}^nY_{ij}x_{ij}\right) \] has formulas of size
$O(n^3)$, and $\Perm_n(\tilde{x}) = \sum_{Y \in \{0,1\}^{n\times n}}
g_n(\tilde{x},Y)$, so $(\Perm_n)$ is in \VNP.

So we have some families in \VP\ (even \VF), and some in \VNP\ but
maybe not in \VP. How do we compare families? For comparing languages,
we have many-one reductions and Turing reductions -- what is the
algebraic analogue? Valiant proposed projections, a most restrictive
kind of reduction when dealing with Boolean classes, but completely
natural in the algebraic context.  We say that $g\in \FieldF[y_1,
  \ldots , y_m]$ is a projection of $f\in \FieldF[x_1, \ldots , x_n]$
if $g$ can be obtained from $f$ by 
substituting a value in $\FieldF \cup Y$ for each variable in $X$.
(For instance, if $f=x_1x_2 + x_3x_4$, then the following are all
projections of $f$: $y_1+y_2$, $y_1y_2+5$, $y_1y_2+y_2y_3$, $2y^2$. But
$y_1^2y_2$, $y_1 + y_2 + y_3$ are not, because a projection cannot
increase the degree or number of monomials.)  Further, we say that a
family $(g_n)$ is a $p$-projection of a family $(f_n)$ if each $g_n$
is a projection of some $f_m$ for an $m$ not too far from $n$. That
is, there is a polynomially bounded function $t$, and each $g_n$ is a
projection of $f_{t(n)}$. If we allow $t$ to be quasi-polynomially
bounded, we obtain $qp$-projections.  

Using these notions of reductions, we have the usual notions of
hardness and completeness for algebraic classes. Here's what Valiant
showed: 
\begin{enumerate}
\item $(\Det_n)$ is hard for \VF\ under $p$-projections (and is known
  to be in \VP). 
\item $(\Det_n)$ is complete for the class of quasi-polynomial size
  formulas \VQF\ under $qp$-projections. 
\item Over fields with characteristic other than 2,  $(\Perm_n)$ is
  complete for \VNP\ under $p$-projections. Over fields of
  characteristic 2, $\Perm_n$ equals $\Det_n$ and hence  is in
  \VP\ and \VQF. 
\item Polynomial families associated with a number of \NP-complete
  languages are complete for \VNP\ under $p$-projections. 
\end{enumerate}
The first two follow from a proof that a polynomial computed by a size
$s$ formula is a projection of $\Det_{s+2}$. (It uses the
combinatorial definition of determinant. as the signed weighted sum of
cycle covers in an associated graph.)  The hardness of $(\Perm_n)$ for
\VNP\ mirrors the hardness of the Boolean permanent for the counting
class $\#P$. As in the case of the upper bound, additional care is
needed to take into account non-access to an input instance and fully
symbolic computations; in particular, the proof requires a
multiplicative inverse of $2$ and hence fails over fields of
characteristic 2. See
\cite{valiant-stoc79a,bcs-book97,burgisser-book00} for various
versions of these proofs. See \cite{Blaser13} for a simplified
gadget construction. 

\section{The current status}
\seclab{current} We now know much more about the classes \VF, \VP,
\VQP, \VNP\ defined above, and about other similarly defined classes.
Let's review these results one by one. 

Say that a family of polynomials $(f_n)$ is a $p$-family if the number
of variables in $f_n$ and the degree of $F$ are polynomially bounded
functions of $n$. We only consider $p$-families.

Recall that $\VP$ consists of $p$-families with polynomial-sized
circuits. Also note that algorithmically, circuit size roughly
corresponds to number of processors needed in a parallel algorithm
(associate one processor per gate), while circuit depth -- the length
of a longest path from the output node to an input node -- corresponds
to parallel time.

A clever construction due to Hyafil \cite{Hyafil79} shows that any
polynomial of degree $D$ in $M$ variables, computable by a circuit of
size $t$, can be computed in parallel time $O(\log D \times
\log(D^2t+M) )$. This is a depth-reduction of the circuit, and
generalises Csanky's result which was specifically tailored for the
determinant. Further, this algorithm has parallel multiplicative depth
only $O(\log D)$; that is, any root-to-leaf path goes through at most
$O(\log D)$ multiplication nodes.  This is worth noting since
multiplication seems a 
more costly operation than addition or subtraction.  Unfortunately,
the resulting circuit, while shallow and depth-reduced, is rather
large, roughly $t^{\log D}$. Applying this construction would take us
from polynomial-size circuits to shallow quasi-polynomial size
circuits. Soon after this, an improved construction was presented by
Valiant, Skyum, Berkowitz and Rackoff \cite{VSBR83}; they achieved the
same depth-reduction (and also $O(\log D)$ multiplicative depth) with
size polynomial in $tD$. In particular, applying this construction to
a circuit family $(C_n)$ witnessing that a polynomial family $(f_n)$
is in \VP, we see that $(f_n)$ is in $\VSAC^1 \subseteq \VNC^2$.  

Wait, what exactly are these new classes? Again, we can think of them
as analogues of Boolean classes.  The Boolean circuit class \NC$^i$
has circuits of polynomial size and $O(\log^i n)$ depth. The class
\SAC$^i$ is similarly defined, polynomial size, $O(\log^i n)$
$\wedge$-depth, and negations only at inputs. That is, if $\vee$ nodes
are allowed to have unbounded in-degree, but $\wedge$ nodes must have
in-degree 2, then these circuits have depth $O(\log^i n)$. (Hence the
name \SAC, for semi-unbounded alternation.)  Clearly, $\NC^i
\subseteq \SAC^i \subseteq \NC^{i+1}$. Now define the classes \VNC$^i$
and \VSAC$^i$ as algebraic analogues of these, with $\times$ and $+$
playing the roles of $\wedge$ and $\vee$ respectively. In the Boolean
world, we know that $\NC^1 \subseteq \SAC^1 \subseteq \NC^2 \subseteq
\ldots \subseteq \NC \subseteq \Ptime$. In the algebraic world,
however, $\VNC^1 \subseteq \VSAC^1 = \VNC^2 = \ldots = \VNC = \VP$.

An important consequence of the depth reduction result of
\cite{VSBR83} is that the $(\Det_n) \in \VQF$ result generalises to all
of \VP; $\VP \subseteq \VQF$.  Another important consequence is that
at quasi-polynomial size, formulas are as powerful as circuits;
\VQF\ equals \VQP. Such an equivalence is not known for $p$-families
at polynomial size. (It holds at exponential size, because polynomials
in any $p$-family have only exponentially many monomials. An explicit
sum-of-monomials expression gives an exponential sized formula.)

Even before the results of \cite{Hyafil79,VSBR83}, 
Brent \cite{Brent} had shown that depth-reduction is possible
for \VF. Any formula $F$ can be rebalanced by identifying in it a
suitably chosen node $N$ and rewriting $F$ as a linear form in $N$,
say $AN+B$. If $N$ is properly chosen, then the polynomials $A$ and
$B$ are computed by small sub-formulas (size at most half of $F$) of
$F$, and can be recursively rebalanced. The appropriate $N$ is
identified by using the tree separator lemma. This process yields a
$O(\log \Size(F))$ depth formula. Thus we conclude $\VF=\VNC^1$.

The depth reduction for \VP\ from \cite{VSBR83} proceeds similarly,
but works on ``proof-trees'' or {\em parse trees}. Unfolding a circuit
into a formula by systematically duplicating reused nodes may yield an
exponential sized formula (recall the example $X^{2^n}$.) Let us
nonetheless do so. Now, a minimal sub-formula that includes the output
node, both children of an included $\times$ node, and exactly one
child of an included $+$ node, computes a potential monomial whose
degree is the number of leaf nodes in the sub-formula. Call such a
sub-formula a proof tree. For a circuit computing a $p$-family of
polynomials, we can ignore proof trees of super-polynomial size. For
each polynomial-sized proof tree, the balancing technique described
above should work. The catch is, there can be too many proof trees
(there can be exponentially many monomials), and each proof tree could
require cutting at a different node. The clever twist is the
following: in the formula depth-reduction, $A$ can be computed
recursively because it is the partial derivative of $F$ with respect
to $N$. If $F$ is now a circuit rather than a formula, then $F$ may
not be linear in $N$, so computing the partial derivative will not
help.  But if $N$ is chosen to have degree more than half the degree
of $F$, then this is indeed the case. So the algorithm of
\cite{VSBR83} computes, for each pair of nodes $N,N'$, a new
polynomial $F(N,N')$; these polynomial are recursively constructed,
and whenever $2\Degree(N)>\Degree(N')$, $F(N,N')$ equals the partial
derivative of $N'$ with respect to $N$. Putting this together
carefully gives the depth-required circuit. For details, see
\cite{VSBR83} itself. Also see \cite{ajmv} and \cite{vollmertext} for
{\em uniform} versions, where the task of describing the
depth-reduced circuit given the original circuit is achieved
using limited computational resources.

A couple of things slipped by almost unnoticed. We know what is meant
by the degree of a polynomial, but what do we mean by
$\Degree(N)$?  This should be the degree of the polynomial computed at
the node $N$, and indeed \cite{VSBR83} use \Degree\ in this sense. But
the uniform versions cannot do so, because computing the degree of a
specified node in a given circuit is a completely non-trivial task!
See the discussion about DegreeSLP in
\cite{ABKM09,kayal-eccc10}. Fortunately, we 
can equally easily work with an upper bound on the degree of each
node. And an upper bound $u(N)$ on the degree at each node $N$ is easy
to obtain: $u(N)=1$ if $N$ is a leaf, $u(N_1+N_2) =
\max\{u(N_1),u(N_2)\}$, $u(N_1\times N_2) = u(N_1) + u(N_2)$. This
upper bound is referred to as the complete formal degree of the
circuit (as opposed to the degree of the polynomial it
computes). However, just because the output node of $C$ computes a
polynomial of degree $d$, this does not imply that each node computes
a polynomial of degree at most $d$. Higher degree monomials may get
computed along the way, and get cancelled finally. Is it necessary, in
terms of efficiency, to compute them?  No! If $C$ is of size $s$ and
computes a polynomial $f$ of degree $d$, then we can construct a
circuit $C'$ of size $O(sd^2)$ computing the same polynomial and with
each node computing a polynomial of degree at most $d$: just compute
the homogeneous parts of $f$ separately in the obvious way. Now $C'$
will have complete formal degree $O(d^3s)$. (See
\cite{MP08} for details.) Thus we could have 
defined \VP\ in terms of circuits of polynomial size and polynomially
bounded complete formal degree as well. 

There is a much simpler proof of the fact that \VP\ is contained in
\VNC. This proof yields a weaker upper bound of \VSAC$^2$ rather than
VSAC$^1$, but is still beautiful, and is still enough to conclude that
$\VP\subseteq \VQF$. I first saw this proof in a survey talk by Pascal
Koiran at Dagstuhl \cite{Koiran-dagstuhl10}, and I wish I had come up
with it myself! Let $(f_n)$ be in \VP, as witnessed by a circuit
family $(C_n)$ with complete formal degree bounded by $(d_n)$. To
depth-reduce $C_n$, partition the nodes into $1+\lceil\log d_n\rceil$
parts; part $k$ has nodes with formal degree in
$[2^{k-1},2^k)$. Treating the polynomials from parts $i < k$ as
  variables, the nodes in part $k$ form a {\em skew} circuit, where
  each $\times$ node has at most one child that is not an input
  node. (Multiplying two nodes both in part $k$ would create high
  degree, giving rise to a node in part $k+1$.)  Now, skew circuits
  can be depth-reduced to \VSAC$^1$ rather easily, using a
  divide-and-conquer argument dating back to Savitch
  \cite{savitch70}. Doing this separately for each part gives a
  \VSAC$^2$ circuit.

We just introduced a new kind of circuit there: skew circuits. Are
they as powerful as general circuits? We do not know! Let's define
\VP$_{skew}$; $p$-families of polynomials computed by polynomial-sized
skew circuits. It turns out this is a great class to study, because it
{\em exactly} characterises the complexity of the determinant. Recall
what we have already seen; $(\Det_n)$ is hard for $\VF=\VNC^1$ and is
in \VP. The upper bound proof from \cite{mahvin} actually gives a skew
circuit of size $O(n^4)$, but skew circuit constructions were known
much earlier: in \cite{ve92}, Venkateswaran first defined Boolean
skew circuits to capture nondeterministic circuits, and subsequently
many authors independently extended that study to arithmetic rings,
\cite{dam91,toda.circ,vin91,val92}. And the lower bound proof from
\cite{valiant-stoc79a} shows that polynomials computed by skew
circuits are $p$-projections of the determinant, though it is not
stated this way. Valiant showed that a formula can be converted to a
certain kind of graph that we nowadays call an algebraic branching
program or ABP (more about this below), and that polynomials computed
by ABPs are $p$-projections of $(\Det_n)$. And we now know that ABPs
are essentially skew circuits.

Time to define ABPs. These are directed acyclic graphs, with a
designated source node $s$ and a designated target sink node $t$
(sometimes there may be multiple target nodes), and with edges labeled
from $\FieldF \cup X$ (similar to input nodes in a circuit). For any
directed path $\rho$, the weight of $\rho$ is the product of the
labels of the edges on $\rho$. The polynomial $p_v$ computed at a node
$v$ is the sum of the weights of all directed $sv$ paths. The
polynomial computed by the ABP is just $p_t$. Families computed by
polynomial-size ABPs form the class \VBP. (In some parts of the
literature, edge labels are allowed to be linear forms in $X$. This
does not significantly change the properties of ABPs as we discuss
here. We'll stick to the convention that labels are in $\FieldF \cup
X$.)

So why are ABPs and skew circuits essentially the same? ABPs to skew
circuits: clearly, $p_s=1$, and for any other source node (in-degree
0) $s'$, $p_{s'}=0$. Look at an edge $u \rightarrow v$ of the ABP with
label $\ell$. Then $p_v$ has a contribution from $p_u \times
\ell$. Summing this over all incoming edges at $v$ gives a small
circuit computing $p_v$ from previously computed values, and this
circuit is skew.  For the reverse simulation, reverse this
construction: (1)~introduce a source node $s$, (2)~for each input node
$u$ labeled $\ell$, add an edge $s\rightarrow u$ labeled $\ell$,
(3)~for each node $v=u + u'$, create edges $u\rightarrow v$ and
$u\rightarrow v$ labeled 1, and (4)~for each node $v=u\times \ell$,
create an edge $u\rightarrow v$ labeled $\ell$.

So now we can add to the list of results at the end of \secref{orig}:
$(\Det_n)$ is complete for 
$\VBP=\VP_{skew}$ under $p$-projections. 

In fact, we can add more. What makes the simulation from skew circuits
to ABPs possible is the fact that at each $\times$ gate, one argument
is {\em easy}. Toda \cite{toda.circ} took this argument further -- it
is enough if one argument is independent of the rest of the
circuit. That is, for each $\times$ node $\alpha = \beta \times
\gamma$, the entire sub-circuit rooted at either $\beta$ or $\gamma$
has no connection to the rest of the circuit except via this edge to
$\alpha$. (Equivalently, one of the edges into $\alpha$ is a bridge in
the circuit.) Call such circuits {\em weakly skew} circuits. Toda
showed that weakly skew circuits can be converted to skew circuits
with linear size blow up. See also \cite{MP08},
where Malod and Portier made the size bounds in the conversions even
more precise. So now we can say $\VBP=\VP_{skew} = \VP_{ws}$, where
the subscript {\em ws} stands for weakly skew. 

(Note: Neither \cite{toda.circ} not \cite{MP08,Malod-thesis} actually
claimed linear size blow-up. However, their constructions from weakly
skew circuits to ABPs, with the standard conversion from ABPs to skew
circuits, does give linear blowup.  As far as I can see, linear blowup
for weakly-skew to skew circuits was explicitly observed in
\cite{KK08,Jan08,Grenet-thesis}.)

Taking this idea further, Malod and Portier provide a brilliant
characterization of the class \VP. Say that a circuit is {\em
  disjoint} if at every node $\alpha = \beta \circ \gamma$, where
$\circ$ could be $+$ or $\times$, the sub-circuits rooted at $\beta$
and $\gamma$ are disjoint.  This is just a fancy (convoluted?) way of
saying that the circuit is a formula. But now relax this constraint a
bit. Say that a circuit is {\em multiplicatively disjoint} or MD if at
every $\times$ node $\alpha = \beta \times \gamma$, the sub-circuits
rooted at $\beta$ and $\gamma$ are disjoint. No restrictions apply to
$+$ nodes. Like formulas, MD circuits of size $s$ have complete formal
degree bounded by $s$. But the MD restriction seems to allow more
computation than formulas; for instance, weakly skew circuits are MD,
and so MD circuits can compute $(\Det_n)$ in polynomial size.  Malod
and Portier showed that in fact polynomial-size MD circuits can
compute everything in \VP, but nothing more. That is,
$\VP=\VP_{\mbox{{\scriptsize MD}}}$. While this fact can also be
deduced once we have depth-reduction to $\VSAC^1$, Malod and Portier
give a completely self-contained combinatorial proof which is very
neat. Basically, imagine that each node in the \VP\ circuit is labeled
with its formal degree. Now make multiple copies of each node,
inversely proportional to the formal degree. By carefully deciding
which copies of its children to use to construct a copy of a node,
multiplicative disjointness can be achieved with only polynomial
blow-up in size.

A nice consequence of this characterisation of \VP\ is a simpler proof
of the fact that \VP\ is contained in $\sum \cdot \VF$.  The key
observation used is that a circuit is multiplicatively disjoint
exactly when every proof tree is already a sub-graph of the circuit
(even without any unfolding into a formula). See \cite{MP08} for details.

Before we move on, we note another surprising relation between ABPs
and formulas: \VF\ equals the class of $p$-families computed by
polynomial-size ABPs of constant width.  What is this resource
``width''? Recall that an ABP is a DAG with edges going ``in the
direction from $s$ to $t$''. Suppose we impose a layering
constraint. The nodes of the DAG must be laid out at the vertices of a
rectangular $w\times \ell$ grid, the node $s$ must be at position
$(S,1)$ for some $S\in [w]$, the node $t$ must be at position
$(T,\ell)$ for some $T \in [w]$, and edges can only go across one
layer, from $(i,k)$ to $(j,k+1)$ for some $i,j \in [w]$, $k \in
[\ell-1]$.  Of course, any ABP can be converted to one of this form:
just sub-divide edges when necessary and label the sub-division path
so that its weight is the original edge's label (use lots of 1s). Now
we say that $w$ is the width of the layered ABP and $\ell$ is the
length. A bounded-width branching program family $(B_n)$ is one where
for some absolute constant $c$, each $B_n$ has width at most
$c$. Seems quite a squeeze -- if we view moving from $s$ towards $t$
as an incremental computation, then at each stage we can carry forward
just $c$ intermediate polynomials. We shouldn't be able to do much
this way, right? Wrong! Ben-Or and Cleve \cite{benor.cleve} showed, in
a proof cleverly extending Barrington's famous characterisation
\cite{bar} of $\NC^1$ by Boolean bounded-with branching
programs, that every formula of depth $D$ has an equivalent
bounded-width branching program (that's quite a mouthful; let's agree
to call it \BWBP) of length $4^D$ and width just 3! Since we already
know that formulas can be depth-reduced and \VF\ equals \VNC$^1$, we
see that \VF\ is contained in a class that we can name \VBWBP:
polynomial-sized constant-width ABPs. The converse inclusion is easily
seen to hold, again using a Savitch-style divide-and-conquer. Thus we
have another characterisation of $\VF$.

As a matter of curiosity, one may want to know: is the width-3 upper
bound tight?  Allender and Wang \cite{AW11} recently settled this
question affirmatively: they show that a very simple polynomial cannot
be computed by any width-2 ABP, no matter what the length. On the
other hand, width-3 ABPs are 
universal, since every polynomial family has some formula family
computing it. The question is one of efficiency: which families have
polynomial-size width 3 ABPs?

OK, so we've had a plethora of class definitions, but just a handful
of distinct classes: $\VF=\VP_e=\VNC^1=\VBWBP$, $\VBP=\VP_{skew} =
\VP_{ws}$, $\VP=\VP_{\mbox{{\scriptsize MD}}}$, $\VQF=\VQP$,
$\VNF=\VNP$. 

As stated in \cite{burgisser-book00}, {\bf Valiant's hypothesis} says
that $\VNP \not\subseteq \VP$, and {\bf Valiant's extended hypothesis}
says that $\VNP \not\subseteq \VQP$.  Over fields of characteristic
not equal to 2, these imply: $\Perm_n$ is not a
$p$-projection of $\Det_n$, and $\Perm_n$ is not a $qp$-projection of
$\Det_n$, respectively.

Some miscellaneous results, in no specific order: 
\begin{enumerate}
\item
Let $\SDet_n$ be the polynomial that represents the determinant of a
symmetric $n\times n$ matrix of indeterminates $B_n$. (For instance,
$\SDet_2=x_{11}x_{22}-x_{12}^2$.) Clearly, $(\SDet_n)$ is a
$p$-projection of $(\Det_n)$.  The converse is also almost true. As
shown by Grenet, Kaltofen, Koiran and Portier in \cite{GKKP11}, over
any field of characteristic other than 2, $\Det_n$ is a projection of
$\SDet_{n^3}$. Characteristic 2 is a problem: symmetric matrices
correspond to undirected graphs, so each undirected cycle gives rise
to two directed cycles, and so to get a projection we need division by
2. In characteristic 2, $\Det_n$
itself is provably not a projection of $\SDet_m$ for any $m$; see
\cite{GMT13}. The best that we can currently say in characteristic 2 is that
the squared determinant $(\Det_n)^2$ is a projection of
$\SDet_{2n^3+2}$; this is also shown 
in \cite{GKKP11}. 

\item \VQP\ is also characterized by quasi-polynomial-size weakly skew
  circuits of polynomial degree. (From \cite{VSBR83} it follows that
  $\VQP = \VQF$; hence the above charcacterization. A direct proof
  is presented in \cite{MP08}.) Several natural polynomials are
  complete for this class under $qp$-reductions: the $(\Det_n)$
  family, of course, but also, the trace of iterated matrix product
  and the trace of a matrix power. These families are all complete for
  \VBP\ under $p$-reductions.

\item
While we do not know the exact relationship between\VQP\ and \VNP,
(they both contain \VP), we do know that \VQP\ does not equal either
\VP\ or \VNP. B\"urgisser (\cite{burgisser-book00}, Section 8.2) has
shown that there is an explicit family of polynomials $(f_n)$ in
\VQP\ that is provably not in \VNP, let alone in \VP. This family is
defined as follows: Consider numbers in base $n$. Let $\mu$ range over
all such numbers with $m(n) = \lceil \log n \rceil$ digits.  More
precisely, let $\mu$ range over length-$m(n)$ sequences over the
alphabet $\{ 0, 1,\ldots , n-1\}$, and let $k_n(\mu)$ denote the value
of this sequence, $k_n(\mu)=\sum_{j=1}^{m(n)}\mu_jn^{j-1}$. Define
$f_n$ as:
\[
f_n(x_1,   \ldots , x_{m(n)}) = \sum_{\mu \in \{0,\ldots ,
  n-1\}^{m(n)}} 2^{2^{k_n(\mu)}} \prod_{j=1}^{m(n)} x_j^{\mu_j}
\]
Exploiting the fact that the distinct double exponentials appear as
coefficients in $f_n$, B\"urgisser shows that $f_n$ cannot be in
\VNP. 

Furthermore, using $m(n)=\lceil \log^i n \rceil$ gives a family of
polynomials $f^i$ in \VQP\ with size $O(n^{\log^i n})$ but provably
not in size $O(n^{\log^{i-1} n})$, so within \VQP\ there is a strict
hierarchy.

\item From the $qp$-completeness of $(\Det_n)$ for \VQP, and the
  $p$-completeness of $(\Perm_n)$ for \VNP, it follows that $\VNP
  \subseteq \VQP$ if and only if $(\Perm_n)$ is a $qp$-projection of
  $(\Det_n)$. This is a very long-standing open question. Originally
  the question of whether $(\Det_n)$ and $(\Perm_n)$ are
  $p$-equivalent was posed by P\'olya \cite{polya1913}, who also
  showed that there is no way of expressing the permanent as the
  determinant by only changing the signs of selected entries (except
  for $n=2$; flip the sign of $a_{12}$ to get matrix $B$ with
  $\Det(B)=\Perm(A)$). (I haven't myself seen P\'olya's note, but have
  seen it referred to in various places.)  Marcus and Minc \cite{mm61}
  showed that there is no size-preserving transformation ($\Perm_n$ to
  $\Det_n$), even if we relax the notion of projections to allow
  linear form substitions for each variable. For many years, a linear
  lower bound was the best known ($\Omega(\sqrt{2}n)$ due to
  \cite{vzG87,Cai90,meshulam}), until Mignon and Ressayre 
  \cite{MR04} showed that over the fields of characteristic 0 (eg real
  or complex numbers), 
  even if linear form substitutions are allowed in projections, to
  express $\Perm_n$ as a projection of $\Det_m$, we need $m \ge
  n^2/2$. The same lower bound was obtained for fields of
  characteristic other than 2 by Cai, Chen and Li \cite{CCL10}.  From
  Ryser's work \cite{ryser1963} it follows that
  $\Perm_n$ is a projection of $\Det_{m}$ for some $m < n^22^n$. 
  More recently, Grenet showed \cite{Grenet12} via a very simple
  and neat construction that   $\Perm_n$ is a projection of $\Det_{m}$
  for $m =2^n-1$.  This is the best known so far. 
  Thus  there is a huge gap between the lower and upper bounds on what is
  called the determinantal complexity of the permanent.

\item It is natural to believe that the complexity of a $p$-family
  $(f_n)$ in this framework is closely related to the computational
  complexity of {\em evaluating} $f_n$ for a given instantiation of
  its variables.  In \cite{burgisser-tcs00}, B\"urgisser gave this
  belief a firm footing. Consider a $p$-family $(f_n)$ where $f_n$
  depends on $n$ variables. Define its Boolean part $\BoolP(f)$ as a
  string function mapping $x\in \{0,1\}^n$ to the binary encoding of
  $f_n(x)$. Note that we have considered only Boolean values. Even so,
  evaluation may seem difficult, because the circuits for $(f_n)$ can
  involve arbitrary constants from the field. B\"urgisser showed that
  assuming the generalised Reimann hypothesis GRH, over fields of
  characteristic zero, $\BoolP(\VP)$ has non-uniform multi-output
  $\NC^3$ circuits. Furthermore, assuming GRH, if Valiant's hypothesis
  is false over such a field, then the entire polynomial hierarchy has
  (non-uniform) \NC\ circuits.
\item
  An {\em extreme} depth-reduction result is given by the highly
  influential paper of Agrawal and Vinay \cite{AV08}. To first see the
  context, note that any polynomial in $n$ variables with degree $d$
  has an unbounded fan-in depth-2 circuit of size
  $2^{O(d+d\log\frac{n}{d})}$. (If $d \in \Omega(n)$, then $2^{O(d)}$
    suffices, otherwise the second term in the exponent makes up.)
    This is because we can just explicitly compute all monomials of
    degree at most $d$, and add up the required ones with suitable
    weights.  Now, can we find circuits substantially better than
    this, say even $2^{o(d+d\log\frac{n}{d})}$, if we allow depth to
    be increased a bit? Agrawal and Vinay showed that indeed this is
    possible, even with depth 4, provided there is some circuit (not
    necessarily depth-reduced) of that size to begin with.  The idea
    is extremely simple. Peform the depth-reduction from \cite{VSBR83}
    or \cite{AJMV98}, and ensure with some additional care that degree
    provably drops at $\times$ gates. (The price for this is small: a
    $\times$ gate may have fanin upto 6, instead of 2.) Now, choose a
    horizontal cut in the depth-reduced circuit so that for the
    sub-circuit above it, and for the sub-circuits below it rooted at
    gates on the cut, the ``brute-force'' construction described above
    is small. Obviously there is a trade-off: if the cut is too high
    up, the lower sub-circuits can have large explicit forms, but if
    it is too low down, the upper sub-circuit can have large explicit
    forms. Cut in the right place, and everything works out! 

    Subsequently the extreme depth-reductions have been pushed
    further; see \cite{Koiran12,Tavenas13,GKKS-focs13}. The lower
    bound results of \cite{GKKS-ccc13,FLMS13} show that the
    depth-reduction upper bound from \cite{Tavenas13} is tight and
    cannot be pushed any further.

    This has significant implications for the quest for derandomizing
    algorithms for the well-studied problem ACIT (arithmetic circuit
    identity testing) --- checking if a given circuit computes the
    identically-zero polynomial.  But that is not directly connected
    with this survey. One question it raises here is: what kind of
    extreme depth-reduction can we achieve for \VQP?  Can we stay
    within quasi-polynomial size? 
\end{enumerate}

\section{The syntactic multilinear world}
\seclab{smlin}

Much of the study concerning \VP\ and \VNP\ involves the families
$(\Det_n)$ amd $(\Perm_n)$. The polynomials in both families are {\em
  multilinear}. In principle, to compute a multilinear polynomial via
a circuit, we need never compute intermediate polynomials that are not
multilinear. Let us call such circuits, where the polynomial computed
at each node is multilinear, {\em multilinear circuits}. However,
often it is the case that allowing non-multilinear terms at
intermediate stages, and eventually cancelling them out, allows more
efficient computation (smaller circuits). 
This leads to the following
quest: what kind of multilinear $p$-families have efficient
multilinear formulas, or even multilinear circuits, where each
intermediate polynomial is required to be multilinear? Even for the
$(\Det_n)$ family, which we know is multilinear and in \VP, we do not
know of polynomial-size multilinear circuits. That being the case, can
we prove lower bounds?

This question is trickier than it seems at first glance, because given
a circuit, even checking whether it is multilinear is
non-trivial. Fournier, Malod and Mengel \cite{FMM12} recently observed
that checking multilinearity of a given circuit is computationally
equivalent to the well-studied problem ACIT (arithmetic circuit
identity testing) --- checking if a given circuit computes the
identically-zero polynomial.

So we may want a notion of certifiably multilinear circuits. One such
notion is that of {syntactic multilinearity}, SM. A circuit is said to
be syntactically multilinear if at every $\times$ node $\alpha = \beta
\times \gamma$, the sub-circuits rooted at nodes $\beta$ and $\gamma$
operate on disjoint sets of variables. Note that this is much more
restrictive than multiplicative disjointness. But it certifies
multilinearity, since no variable can ever get multiplied by
itself. And syntactic multilinearity is easy to check computationally:
it is violated if there is some node $\alpha = \beta \times \gamma$,
some variable $x$, two input nodes $I,I'$ labeled $x$, and paths from
$I$ to $\beta$ and $I'$ to $\gamma$.

If a family has efficient (polynomial-sized) SM circuits, then it has
efficient multilinear circuits. The converse may not be true. But it
is true if we look at formulas. Given a multilinear formula, identify
an SM violation $\alpha,\beta,\gamma,x$ as above. Then we know by
multilinearity of the polynomial $p(\alpha)$ that $x$ does not appear
in either $p(\beta)$ or $p(\gamma)$. In the appropriate sub-formula,
set all instances of $x$ to 0; the polynomials computed at and above
$\alpha$ remain unchanged. Doing this systematically gives an SM
formula of size no more than the original multilinear formula. 

In the first major breakthrough, Ran Raz \cite{raz-det-perm} showed
that for computation by SM formulas, and hence by multilinear
formulas, both $(\Det_n)$ and $(\Perm_n)$ need size $n^{\Omega(\log
  n)}$.  Clearly, this also means that they are not in SM-\VNC$^1$.  

Since $(\Det_n)$ is in \VP\ and even in \VBP, SM-\VF\ is strictly
weaker than \VBP. But this is hardly a fair comparison: we have
restricted \VF\ to be SM, but not \VBP\ and \VP. Can we say that
SM-\VF\ is strictly weaker than SM-\VBP\ or SM-\VP? We do not know
whether $(\Det_n)$ is in multilinear \VP, let alone SM-\VP, so a
different family is needed as a separating example.  Such an example
was provided soon thereafter, again by Ran Raz \cite{raz-mlin-nc}. He
constructed an explicit polynomial family that is in SM-\VP\ and even
in SM-\VSAC$^1$, and showed that it needs SM-formula size
$n^{\Omega(\log n)}$ and hence is not in SM-\VNC$^1$. Improved lower
bounds for constant-depth circuits and subclasses of formulas were
subsequently obtained by Raz, Shpilka and Yehudayoff (see for instance
\cite{ry-cc09}, \cite{rsy08}).

Let's step back a bit. Why did we say ``in SM-\VP, and even in
SM-\VSAC$^1$``? Aren't \VP\ and \VSAC$^1$ the same? Well, we know that
\VP\ and \VF\ can be depth-reduced. But can we assume that these
depth-reduction tehniques preserve syntactic multilinearity?
Fortunately, they do; Raz and Yehudayoff \cite{RY08-bal} showed that
the depth-reduction of \cite{VSBR83} preserves SM, so indeed \SM-\VP
= \SM-\VSAC$^1$. Similarly, in \cite{JMR12} it is observed that the
formula depth-reduction of \cite{Brent} also does preserves SM, so
\SM-\VF = \SM-\VNC$^1$.

What about other relationhips between the algebraic classes? We had
considered ABPs -- what certifies multilinearity there? It is easy to
see that a read-once restriction, where on each path in the ABP each
variable appears as a label at most once, does so. Let us therefore
use read-once as the definition of syntactic multilinearity in
ABPs. Then, as observed in \cite{JMR12}, the Savitch-style divide-and
conquer argument preserves SM. So does the conversion from formulas to
ABPs, \cite{valiant-stoc79a}. But the conversion from formulas to
width-3 ABPs, \cite{benor.cleve}, does not. In fact, Rao
\cite{rao-thesis} showed that even a significant generalisation of
Ben-Or and Cleve's technique, using polynomially many registers instead
of just 3, cannot preserve syntactic multilinearity. Of course, there
may be other ways of going from SM-\VF\ to SM-\VBWBP, but it could
equally well be that the classes are distinct.

To get back perspective, in the SM world what we have seen so far is: 
\[\SM\text{-}\VBWBP \subseteq \SM\text{-}\VF \subseteq \SM\text{-}\VBP 
\subseteq \SM\text{-}\VP\] As mentioned earlier, Raz
\cite{raz-mlin-nc} showed that the inclusion from \SM-\VF\ to
\SM-\VP\ is proper.  Very recently, this was improved by Dvir, Malod,
Perifel, and Yehudayoff \cite{DMPY12}. They showed that in fact the
inclusion $\SM\text{-}\VF \subseteq \SM\text{-}\VBP$ is strict.
Whether the first and the last inclusion are strict is still open.

The proof of \cite{DMPY12} is a clever adaptation of the original
technique from \cite{raz-mlin-nc}.  Let us briefly examine this.

The central ingredient in Raz's proof is randomly partitioning the
variables and analysing the rank of the resulting partial derivatives
matrix. Consider a polynomial $f$ on $2n$ variables $X=\{x_1, \ldots ,
x_{2n}\}$, and consider a partition of $X$ into equi-sized sets $Y$,
$Z$. Consider a $2^n \times 2^n$ matrix $M_{f}^{Y,Z}$ where rows and
columns are indexed by subsets of $Y$ and $Z$ (equivalently,
multilinear monomials over $Y$ and $Z$ respectively). The entry
$(m_y,m_z)$ is the coefficient of the monomial $m_y \cdot m_z$ in
$f$. Intuitively, if $M_{f}^{Y,Z}$ has high rank, then $f$ should be
hard. But high rank with respect to what partition? Raz showed that if
multilinear $f$ has small SM-formula size, then for at least one
partition $(Y,Z)$ of $X$, $M_{f}^{Y,Z}$ will have low rank. (The
existence of the partition witnessing low rank is proved using the
probabilistic method; choose a partition at random, and analyse the
probability that the resulting matrix has rank exceeding some
threshold.) He also constructed an explicit family $g$ in \SM-\VSAC$^1$
and showed that for every partition $(Y,Z)$ of $X$, $M_{g}^{Y,Z}$ has
high rank; hence $g$ is not in \SM-\VF.

The non-trivial adaptation done in \cite{DMPY12} is to consider not
all partitions, but a fairly small set of what they call
arc-partitions.  They showed that if $f$ is in \SM-\VF, then for at
least one arc-partition $(Y,Z)$ of $X$, $M_{f}^{Y,Z}$ will have low
rank.  They consider an explicit family $g$ in \SM-\VBP\ and show that
for every arc-partition $(Y,Z)$ of $X$, $M_{g}^{Y,Z}$ has high
rank. Hence $g$ is not in \SM-\VF. The low-rank proof is again
probabilistic, but it has a very appealing combinatorial flavour. So
does the very definition of an arc-partition.


\section{More on completeness}
\seclab{complete}

Assume that completeness is defined with respect to
$p$-projections. If a family $(f_n)$ is complete for a class, then
understanding $(f_n)$ better allows us to understand the class
better. If a natural family is complete for a class, then this is
evidence that the class itself is natural.

Valiant started off with a proof that \Perm\ is \VNP-complete. He also
showed that polynomial families associated with a number of
\NP-complete languages are complete for \VNP\ under $p$-projections.
So let us agree that \VNP\ is a natural class. 

What about \VP? The family that naturally contrasts with \Perm\ is
\Det, but \Det\ is not yet known to be complete for \VP\ (unless we
allow $qp$-projections; that is not quite satisfactory). If this turns
out to be the case, it will solve a major open problem, showing that
polynomial-degree polynomial-size circuits are no more powerful than
polynomial-size branching programs \VBP.  \VBP\ seems a natural enough
class, and  \Det\ and many other families are complete for it. 

So what problems are complete for \VP?  One can construct a canonical
family complete for \VP. By canonical, I mean something similar to
saying that 
\[\{\langle M,x,1^t\rangle \mid \mbox{~$M$ is an NDTM that accepts $x$
in $t$ or fewer steps} \}\] is \NP-complete. Undoubtedly true, but it
doesn't give any new intution about what \NP\ is about. In the case of
\VP, the canonical family is not so trivial to construct (but not very
difficult either).  

The first description, with a very general completenes result, appears
in \cite{burgisser-book00} (see section 5.6, Cor 5.32(b)).
B\"urgisser shows that for every $p$-family $h$, the {\em relativized}
classes $\VP^h$ and $\VNP^h$ have complete families with respect to
$p$-projections.  Since $\VP^h=\VP$ and $\VNP^h=\VNP$ whenever $h$
itself is in \VP, this gives families complete for $\VP$ and \VNP\ as
well. (In fact, it shows the existence of \VNP-complete families,
independent of Valiant's original proof.)  These complete families
compute homogeneous components separately, to keep the degree small,
and then add up the required parts. They are constructed by first
defining {\em generic polynomials}, and then defining the appropriate
projection / substitution. The generic polynomials capture the
canonical notion referred to above.

Later, a more direct construction tailored for \VP\ (as opposed to
$\VP^h$ and $\VNP^h$ for all $h$) 
was  described by Ran Raz
\cite{raz-elusive}, and also appears in \cite{SY10-survey}. 
Here the proof of hardness exploits the fact that we can perform
depth-reduction on \VP\ circuits. (This was not needed in
B\"urgisser's proof.)  Roughly, here's how it goes: For each natural
number $N$, consider a circuit $C_N$ with nodes arranged in $2\log N +
1$ layers numberd $0,1,\ldots ,2\log N$.  All even layers have exactly
$N$ nodes, and compute polynomials $g_{i,j}$ where $i$ is the layer
number, $ j \in [N]$. Odd layers are used to build these
polynomials. At layer 0, the polynomials are just distinct variables,
$g_{0.j}= x_j$. At higher layers, we have an inductive definition:
$g_{i+1,j} = \sum_{k,\ell\in [N]} g_{i,k} \cdot g_{i,\ell} \cdot
y_{i,j,k,\ell}$, where the $y_{i,j,k,\ell}$ are new variables. Thus
the nodes at the odd layers are the fanin-3 $\times$ nodes, and nodes
at even layers (other than the 0 layer ) are $+$ nodes with large
fanin. (We can reduce the fanins to 2 later; it won't change the
polynomial computed.) The polynomial computed by this circuit at
$g_{2\log N,1}$ is $p_N$. The total number of variables is $O(N^3 \log
N)$, and the circuit is also of size $O(N^3 \log N)$. The degree of
$p_N$ is $2N-1$.  So $(p_N)$ is in \VP. Why is it \VP-hard?  Take any
family $(f_n)$ in \VP. By the depth-reduction of \cite{VSBR83}, it can
be computed in $\VSAC^1$. The $\VSAC^1$ circuit $D_n$ can be
normalised to have alternating $+$ and $\times$ nodes, with all
$\times$ nodes having fanin 2, and all leaves at the same
depth. Choose $N$ at least as large as $\min\{\Size(D_n),
2^{\Depth(D_n)}\}$, and also at least as large as the number of
variables in $C_n$. Now, the computation of $D_n$ can be embedded into
$C_N$: Choose the right number of $+$ nodes at each even layer, and by
carefully assigning 0,1 values to the $y$ variables, ensure that they
compute the required combinations of polynomials from the previous
even layer.

The circuits described above are called {\em universal circuits} in
\cite{SY10-survey}, because every circuit is a projection of the
universal circuit of appropriate size. And if we start with
\VP\ circuits, the projections are $p$-projections. 

So now we know that \VP\ has complete families under $p$-projections
as well. But generic polynomials, universal circuits, and the
polynomials they compute, are rather artifical. Are there other
families that are defined independent of circuits and are
\VP-complete? Actually, we know very few.
Recently, Stefan Mengel \cite{mengel11} made further progress here,
considering polynomial families associated with constraint satsfaction
problems CSPs. (This builds on earlier work by Briquel, Koiran, Meer
\cite{BK09,BKM11}, though they did not explicitly look for
\VP-completeness.) Let's first review what CSPs are. Think of them as
generalising CNF-SAT. In CNF-SAT, each clause forbids one assignment
to the variables in it. (eg the clause $x_1 \vee \overline{x_3}$
forbids $x_1=0,x_3=1$.)  In a CSP, variables can take values from a
larger domain, not necessarily 0,1. Each constraint is like a clause;
it has a set of variables, and it forbids certain combinations of
assignments to these variables. (eg on domain $\{a,b,c\}$ a constraint
on $x_1,x_2$ could say that $x_1 \neq x_2$. That is, assignments
$aa,bb,cc$ are forbidden, the other 6 assignments satisfy this
constraint.)  As in SAT, we look for assignments satisfying all
constraints. If the domain has size 2, the CSP is Boolean. If each
contraint involves 2 (or less) variables, the CSP is binary. As usual,
consider not just a CSP but a family of CSPs $(\Phi_n)$, where
$\Phi_n$ has domain $D_n$. For tractability, we will require that
the CSP is $p$-bounded; that is, the CSP has bounded arity (for some
fixed constant $c$, each constraint in every $\Phi_n$ looks at no more
than $c$ variables), and it has polynomial sized domains (in $\Phi_n$,
the variables take values from a set $D_n$, where the size of $D_n$ is
$p$-bounded). Now associate with each such CSP $(\Phi_n)$ a polynomial
family $(Q_n=Q(\Phi_n))$, where $Q_n$ is on the variable set $\{ X_d
\mid d \in D_n\}$ and is defined as follows:
\begin{eqnarray*}
Q(\Phi_n) 
&=& \sum_{a:\Var(\Phi_n) \rightarrow D_n} 
  [\mbox{$a$ satisfies $\Phi_n$}] \prod_{x\in\Var(\Phi_n)} X_{a(x)}
\\
&=& \sum_{a:\Var(\Phi_n) \rightarrow D_n}  [\mbox{$a$ satisfies $\Phi_n$}] 
\prod_{d\in D_n} X_d^{|a^{-1}(d)|}
\end{eqnarray*}
(Recall, $[S]$ is Boolean, 1 if and only if statement $S$ is true.)
Mengel has this wonderful characterization of the complexity of the
family $(Q_n)$. The characterization involves associating with the CSP
a graph $G$; this graph has a vertex for each variable and an edge
between two variables if they occur simultaneously in some
constraint. Now the treewidth and pathwidth of the graph (these
parameters describe roughly how tree-like or path-like the graph is,
if we can consider blobs of vertices. The smaller the blobs, the
better the similarity. See \cite{Bodlaender-tcs98} for definitions and
an overview.) relate to the complexity.  It also involves an
assignment bound: a CSP is $c$-assignment-bounded if for each
constraint $\varphi$ and each variable $x$ in the constraint, the
number of distinct values possible for $x$ in assignments satisfying
$\varphi$ is bounded by $c$, even though the domain may be much
larger. This seems like a strong condition, but recall that Boolean
CSPs are by definition 2-assignment-bounded.

Enough of definitions! Here's what Mengel shows:
\begin{enumerate}
\item For each $p$-bounded CSP $(\Phi_n)$, $(Q(\Phi_n))$ is in
  \VNP. Every family $(f_n)$ in \VNP\ is a $p$-projection of
  $(Q(\Phi_n))$ for some $p$-bounded $(\Phi_n)$.
\item For each $p$-bounded CSP $(\Phi_n)$ where $G_n$ has bounded
  treewidth, $(Q(\Phi_n))$ is in \VP. Every family $(f_n)$ in
  \VP\ is a $p$-projection of $(Q(\Phi_n))$ for some $p$-bounded
  binary $(\Phi_n)$ where $G$ is a tree (treewidth 1).
\item For each $p$-bounded CSP $(\Phi_n)$ where $G_n$ has bounded
  pathwidth, $(Q(\Phi_n))$ is in \VBP. Every family $(f_n)$ in
  \VBP\ is a $p$-projection of $(Q(\Phi_n))$ for some $p$-bounded
  binary $(\Phi_n)$ where $G$ is a path (pathwidth 1).
\item For each $p$-bounded $c$-assignment-bounded CSP $(\Phi_n)$ where
  $G_n$ has bounded treewidth, $(Q(\Phi_n))$ is in \VF. Every family
  $(f_n)$ in \VF\ is a $p$-projection of $(Q(\Phi_n))$ for some
  $p$-bounded 2-assignment-bounded binary $(\Phi_n)$ where $G$ has
  pathwidth at most 26.
\end{enumerate}
The hardness proofs involve looking at the structure of parse trees
for \VP, witnessing paths for \VBP. 

Note that as stated, this falls slightly short of providing a single
complete family for \VP.  However, applying the hardness reduction
from universal circuits will yield a single CSP family that is
\VP-complete.  To the best of my knowledge, this is the first instance
of a \VP-hardness result for a family defined (almost) independent of
circuits.

All the above results require that the CSP has bounded
arity. Unbounded arity seems to immediately give rise to
intractability.  If arity is unconstrained, can other types of
restrictions still result in families in \VP? For further progress in this
direction, see \cite{DM11,CDM13}. 

\section{Computing integers}
\seclab{tau} 
The questions concerning algebraic complexity classes are
closely connected to another very intriguing question.  Let $N>1$ be
any natural number. Suppose we want to build up $N$ from 1, using only
$+$, $-$ and $\times$. The most naive way of doing this would be
$N=1+1+\ldots +1$. But depending on $N$ there can be many other
ways. Which is the {\em most efficient} way? That is, which way uses
the least number of $+$ or $\times$ operations? To do anything
non-trivial, we must use $+$ at least once, and the first time we use
it we will generate 2.  So let us not even count this mandatory
$+$. How many more operations are needed? 

We can state this as a question about circuits. Each way of building
up $N$ is an arithmetic circuit, or a straight-line program (SLP),
that uses no constants other than 1 and 2. Let us denote by $\tau(N)$
the size of the smallest such circuit computing $N$. (This is the
$\tau$ complexity of $N$).  By definition, $\tau(1)=\tau(2)=0$, and
for all $N>2$, $\tau(N) >0$. Algorithms for computing $N$ give upper
bounds on $\tau(N)$. For instance, to compute $N=2^k$, here's an SLP:
$g_0=2$, $g_{i+1} = 2\times g_{i}$ for $0 \le i \le k-2$. Clearly,
$g_i$ computes $2^{i+1}$, so $\tau(2^k) \le k-1$. But I'm sure you can
already see better ways of doing this. From the circuit viewpoint, an
explanation of why this is not the best is that the circuit
corresponding to this SLP is skew. Surely we should be able to use
non-skew gates and compute large numbers faster. Here's another SLP
that computes big numbers fast: $f_0=2$, $f_{i+1} = f_i \times f_{i}$
for $0 \le i \le \ell-1$. Clearly, $f_i$ computes $2^{2^i}$, so
$\tau(2^{2^\ell}) \le \ell$, a much better bound than the earlier
$2^{\ell}-1$ at least for numbers of this form. Note that the way we
used non-skewness, we produced a circuit with exponential formal
degree (the degree at $f_i$ is $2^i$), but we're not worried about
that for now. Now, using these compact circuits for $2^{2^\ell}$, we
can build a better circuit for $2^k$ by just using the binary
expansion of $k$: $k = \sum_{i=0}^t b_i2^i$, where $t = \lfloor \log k
\rfloor$ and $b_t = 1$. So $2^k = 2^{\sum_{i=0}^t b_i2^i} =
\prod_{i=0}^t 2^{b_i\times 2^i} = \prod_{i: b_i=1}2^{2^i}$. Compute
all the double powers using $t$ operations, and then multiply the
required ones using at most $t$ operations. Overall, $\tau(2^k) \le 2t
= 2 \lfloor \log k \rfloor$.

We can use the same binary expansion idea to compute any $N$, not just
a power of 2. Compute all powers of 2 upto $\log N$, and add the
required ones. This shows that for all $N$, $\tau(N) \le 2\lfloor \log
N\rfloor -1$. 

So far we have not used any subtractions. But they can be very useful
too.  For instance, $\tau(2^{2^\ell}-1) \le \ell + 1$; compute
$2^{2^\ell}$ and subtract 1. 

What about a lower bound? We can actually formalise the
intuition that the exponential degree circuits we saw above for
$2^{2^\ell}$ produce the largest possible number in that size. Hence,
for any $N$, $\tau(N) \ge \log \log N$.   

In particular, $\tau(2^{2^\ell}) = \ell$. That sounds impressive -- we
know the exact value of $\tau$ for $2^{2^\ell}$. 
But essentially just for that;
for all other numbers, we still seem to have a pretty large gap. If
$N=2^k$, then $\log \log N \le \tau(N) \le 2 \lfloor \log k \rfloor =
2 \lfloor \log \log N \rfloor$, so we know $\tau(N)$ within a factor
of 2. But for general $N$, all we know is $\log \log N \le \tau(N) \le
2\lfloor \log N\rfloor -1$. How can we reduce this gap? An obvious
search for an efficient way where the last operation is $+$ or $-$ is to
express $N$ as $M\pm k$, compute $M$, compute $k=\pm(N -M)$, and combine, and to
choose $M$ that minimizes $\tau(M)+\tau(k)+1$. (A similar approach
can be used for factors of $N$ and a $\times$ as the last operation.)
But in computing $M$ and $\pm(N- M)$ (or $N/M$), the complexity may be {\em
  sub-additive} since we can reuse intermediate numbers from the
program for $M$ while computing $\pm(N-M)$ or $N/M$. (We are looking for
circuits, not formulas.)  It is identifying the extent of this reuse
that is a challenge.

Similar to Shannon's bound for functions and circuits (most functions
require exponential sized circuits), de Melo and Svaiter \cite{ms-ams96}
showed that most numbers $N$ have $\tau(N)$ closer to the upper
bound. They showed that for every $\epsilon >0$, most $N$ satisfy
$\tau(N) \ge \frac{\log N}{(\log \log N)^{1+\epsilon}}$. Moreira
\cite{moreira-ams97} improved this by showing that this holds even for
$\epsilon=0$. (He also showed that for all $\epsilon > 0$, there is an
$N_\epsilon$ such that for all $N \ge N_\epsilon$, $\tau(N) \le
\frac{(1+\epsilon)\log N}{(\log \log N)}$.) And yet, showing such lower
  bounds for specific numbers seems quite hard -- the classic
  ``searching for hay in a haystack'' paradox.

Let's move over from individual numbers to sequences of numbers. Let
$(a_n)_{n\ge 1}$ be some sequence of natural numbers.  When can we say
that the sequence is easy to compute? Each number in the sequence
should be ``easy'' relative to its position in the sequence. That is,
the sequence $(b_n)$, where $b_n = \tau(a_n)$, should not grow very
fast. One possible definition is that $b_n$ should be polynomially
bounded in $n$.  For instance, for $a_n=2^{2^n}$, we know that
$b_n=n$. Is that not moderate growth? Not really.  Consider a function
that maps a position $n$ to not just the number $\tau(a_n)=b_n$ but to
an SLP of size $b_n$ computing $a_n$. For the sequence $(2^{2^n})$,
this function takes an input $n$ represented in $\Theta(\log n)$ bits,
and outputs a circuit of size $n$, that is, exponential in the size of
the input. That's not moderate growth!

OK, so let's say that a sequence $(a_n)$ is easy to compute if for
some polynomial $p(.)$, for each $n$, $\tau(a_n) \le p(\log n)$, and
otherwise it is hard to compute. We've set up this definition so that
$(2^{2^n})$ is hard to compute, while the sequences $(n)$, $(2^n)$ are
easy to compute. Makes sense? Now let's ask, what other sequences are
easy? And what sequences are hard?

A sequence with famously open status is $(n!)$.  The completely naive
SLP that constructs the first $n$ numbers with $n-2$ increments and
then multiplies them shows that $\tau(n!) \le 2n-4$. But can this be
improved significantly? Or is this sequence hard?  The best we know is
that $\tau(n!) \in O(\sqrt{n} \log^2 n)$; see \cite{bcs-book97}. Here
is the interesting connection to algebraic circuit
complexity. Building on a sequence of constructions by Cheng
\cite{Cheng04} and Koiran \cite{Koiran05}, B\"urgisser \cite{Bur-cc09}
showed that if $(n!)$ is hard to compute, then any algebraic circuit
for the $(\Perm_n)$ family that uses only the constants $-1,0,1$ 
must be of
superpolynomial size. If we can't even compute the numbers $n!$
easily, then we cannot compute the polynomials $\Perm_n$ efficiently,
unless we allow the use of constants that cannot themselves built
up efficiently. 

Analogous to the $\tau$ complexity of natural numbers, we can define
the $\tau$ complexity of polynomial families. Let $\tau(f)$ denote the
size of the smallest algebraic circuit using only the constants
$-1,0,1$ -- call such a circuit {\em constant-free} -- and computing
$f$. We say that the family $(f_n)$ has polynomially bounded $\tau$
complexity if for some polynomial $p(n)$, and for each $n$, $\tau(f_n)
\le p(n)$. B\"urgisser's result can now be stated as: if
$\tau(\Perm_n)$ is polynomially bounded, then $(n!)$ is easy to
compute.

Let's examine this a bit closely. Why do we state the hypothesis as
``$\tau(\Perm_n)$ is polynomial''? Is this not equivalent to saying
$(\Perm_n)$ is in \VP, and hence $\VNP=\VP$? Actually, it may not be
equivalent. It is possible that $(\Perm_n)$ has polynomial-sized
circuits but no polynomial-sized constant-free circuits.  Conceivably,
using other constants in intermediate computation and then cancelling
them out could help. Recall that the proof of \VNP-hardness of
$(\Perm_n)$ uses constants other than $-1,0,1$; $1/2$ is needed. (As
another example, recall how in showing that $\Det_n$ is a projection
of $\SDet_n$, we needed the constant $1/2$, even though all
coefficients in $\Det_n$ are $-1,0,1$.)  So we can define a subclass
of \VP: families with constant-free circuits of polynomial size.

What can we say about such a subclass?  As described above,
B\"urgisser has shown that if this subclass contains $(\Perm_n)$, then
$(n!)$ is easy to compute. Under the same hypothesis, he also shows
that the sequences $\lfloor 2^n e \rfloor$, $\lfloor (3/2)^n \rfloor$
and $\lfloor 2^n \sqrt{2} \rfloor$ are easy to compute.

Malod \cite{Malod-thesis} observed that unlike in the case of \VP, for
constant-free circuits we may not be able to bound complete formal
degree.  For \VP, if the polynomial computed by a circuit of size $s$ had degree
$d$, we could find an equivalent circuit with  formal degree
$d$, and another with complete formal degree $O(d^3s)$, 
with only polynomial  blow up in size. Not so if constants aren't freely
available!  Consider the polynomial family $f_n=2^{2^n}(x_1 + \ldots +
x_n)$. With arbitrary constants, we have a circuit of size $n$. With
only $-1,0,1$, we have a circuit of size $2n+1$: build $2$, build
$2^{2^n}$, build the linear form, multiply. But this circuit has
exponential formal degree, and in fact, using only the constants
$-1,0,1$, any circuit must have exponential formal degree to build up
$2^{2^n}$. So this polynomial is in \VP, it has constant-free circuits
of polynomial size, but it does not have constant-free polynomial-size
circuits with 
polynomially-bounded complete formal degree.

This leads to a definition of a further subclass $\VP^0$, first
defined in \cite{Malod-thesis}: polynomial families computed by
constant-free circuits with polynomially bounded complete formal
degree.  Define $\VNP^0$ analogous to $\VNP$ as $\sum \cdot
\VP^0$. Check back; our proof that $(\Perm_n)$ is in \VNP\ also shows
that $(\Perm_n)$ is in $\VNP^0$.

The hypothesis $(\Perm_n) \in \VP^0$ is stronger than saying that
$\tau(\Perm_n)$ is polynomially bounded. What does it imply? Can it
lead to more sequences being easier to compute? Firstly, note that
$(\Perm_n) \in \VP^0$ does not immediately imply $\VP^0=\VNP^0$.  All
we can say is the following, shown by Koiran \cite{Koiran05}: If
$(\Perm_n)$ is in $\VP^0$, then for every family $(f_n)\in\VNP^0$,
there is some polynomially-bounded function $p(n)$ such that the
family $(2^{p(n)}f_n)$ is in $\VP^0$. That is, a ``shifted'' version
of $f_n$ is in $\VP^0$. The precise shift can be described as follows
-- we know that $f_n$ is a projection of $\Perm_{q(n)}$ for some
polynomially bounded $q(n)$, we assumed that $\Perm_{q(n)}$ can be
computed by a circuit $C_n$ of size and formal degree bounded by a
polynomial function of $n$, we take $p(n)$ to be the formal degree of
$C_n$. Now $C_n$ can be massaged to compute $2^{p(n)}f_n$ instead of
$\Perm_{q(n)}$.

This motivates another variant of easy-to-compute.  Let's say that a
sequence $(a_n)$ of natural numbers is ultimately easy to compute if at
least some shifted version of it is easy to compute. That is, there is
some other integer sequence $A_n$ such that the sequence $a_nA_n$ is
easy to compute.  Note that if $(a_n)$ is not ultimately easy, then
for infinitely many $n$, all non-zero multiples of $a_n$ have large
$\tau$ complexity.  Using this property, under the hypothesis that
$n!$ is not even ultimately easy to compute, we can obtain a
non-trivial derandomization of the Arithmetic-Circuit-Identity-Testing
problem; see the last section of \cite{ABKM09}. Earlier, Koiran showed
in \cite{Koiran05} that if $n!$ is not even ultimately easy to
compute, then we have some separation: either $\VP^0 \neq \VNP^0$, or
$\Ptime \neq \PSPACE$. This is curious: we have a consequence
involving Boolean classes as well. But it should not be so
surprising. $\VP^0$ and $\VNP^0$ are computed by (sums of)
constant-free poly-formal-degree algebraic circuits, and these are the
arithmetic circuits that arise when we consider counting classes like
$\#\Ptime$ that count accepting paths of Turing machines. This does
not mean that $\VNP^0=\#\Ptime$; the former is a collection of
polynomial families whereas the latter is a collection of functions
from strings to whole numbers. But the complexity of evaluating
polynomial families in the former collection, at Boolean arguments, is
closely related to what the latter collection refers to. Koiran's
proof actually shows the contrapositive: he first shows that if $\VP^0
= \VNP^0$ and $\Ptime = \PSPACE$, then the sequence $\tau((2^\ell)!)$
is polynomially bounded in $\ell$. So consider instead of each $n!$
the possibly larger factorial $(2^{\ell(n)})!$, where $2^{\ell(n)-1} <
n \le 2^{\ell(n)}$.  Then the sequence $(b_n) = ((2^{\ell(n)})!)$ is easy
to compute, and each $b_n$ is a multiple of $n!$, so $(n!)$ is
ultimately easy to compute.

Since $\Perm_n$ is not known to be complete for $\VNP^0$, what is?  It
turns out that for several other \VNP-complete families, the hardness
proofs use no constants other than $-1,0,1$ and the membership proofs
use circuits with small formal degree; hence these families are
complete for $\VNP^0$ as well. As a concrete example, consider the
Hamilton cycle polynomial family $\HC_n$ defined as follows: Let
distinct variables $x_{i,j}$ label the edges of the complete directed
graph $D_n$. Let $C_n$ denote the set of all directed Hamiltonian
cycles in $D_n$; elements of $C_n$ can be described by cyclic
permutations $\sigma\in S_n$. Then
\[ \HC_n(x_{11}, \ldots , x_{nn}) = \sum_{\sigma \in C_n} \prod x_{i,\sigma(i)}\]
This family is complete for $\VNP^0$; (\cite{Malod-thesis}).

Returning to the question ``What does $(\Perm_n) \in \VP^0$ imply?'';
Koiran \cite{Koiran05} showed that it implies the sequence $\lfloor
2^n \ln 2 \rfloor$ is easy to compute. He also improved the
earlier-mentioned result in two ways, from ``$[(\VP^0=\VNP^0) \wedge
(\Ptime = \PSPACE)] \Rightarrow (n!)$ is ultimately easy to
compute'' to ``$[(\Perm_n \in \VP^0) \wedge (\Ptime = \PSPACE)]
\Rightarrow (n!)$ is easy to compute''.

Under the stronger hypothesis that $\VP^0=\VNP^0$, we can show more
(again due to \cite{Koiran05}).  If $\VP^0=\VNP^0$, then the sequences
$(\sum_{i=1}^{2^n} 2^{i^2-1})$, $\lfloor 2^{2^n} \ln 2 \rfloor$,
$\lfloor 2^{2^n} \ln 3 \rfloor$, $\lfloor 2^{2^n} \pi \rfloor$, all
have polynomially bounded complexity, something that is not yet known
unconditionally.

\section*{Acknowledgements}
I thank Arvind and Manindra for inviting me to contribute to this
volume in honour of Somenath Biswas, a wonderful professional
colleague and friend. 

I thank CEFIPRA for supporting an Indo-French collaboration (project
4702-1); many of my ideas for how to present this survey were
crystallised during my visit to University of Paris-Diderot during
May-June 2012.

I have picked material I find interesting, and have not really
attempted an exhaustive coverage. I apologize in advance to those
whose favourite results I have omitted.

I gratefully acknowledge many insightful discussions with Eric
Allender, V Arvind, Herv\'e Fournier, Bruno Grenet, Nutan Limaye,
Guillaume Malod, 
Stefan Mengel, Sylvain Perifel, B V Raghavendra Rao, Nitin Saurabh,
Karteek Sreenivasaiah, Srikanth Srinivasan, V Vinay. I thank the
organisers of the Dagstuhl Seminars on Circuits, Logic and Games (Feb
2010) and Computational Counting (Dec 2010) for inviting me and giving
me the opportunity to discuss these topics.  The survey by Pascal
Koiran at the Dagstuhl seminar on Computational Counting in Dec 2010
was particularly helpful.

\bibliographystyle{alpha}
\bibliography{../../master}

\begin{thebibliography}{ABKPM09}

\bibitem[ABKPM09]{ABKM09}
Eric Allender, Peter B{\"u}rgisser, Johan Kjeldgaard-Pedersen, and Peter~Bro
  Miltersen.
\newblock On the complexity of numerical analysis.
\newblock {\em SIAM J. Comput.}, 38(5):1987--2006, 2009.

\bibitem[AJMV98a]{ajmv}
E.~Allender, J.~Jiao, M.~Mahajan, and V.~Vinay.
\newblock Non-commutative arithmetic circuits: depth reduction and size lower
  bounds.
\newblock {\em Theoretical Computer Science}, 209:47--86, 1998.

\bibitem[AJMV98b]{AJMV98}
Eric Allender, Jia Jiao, Meena Mahajan, and V.~Vinay.
\newblock Non-commutative arithmetic circuits: Depth reduction and size lower
  bounds.
\newblock {\em Theor. Comput. Sci.}, 209(1-2):47--86, 1998.

\bibitem[AV08]{AV08}
Manindra Agrawal and V.~Vinay.
\newblock Arithmetic circuits: A chasm at depth four.
\newblock In {\em FOCS}, pages 67--75, 2008.
\newblock See also ECCC TR15-062, 2008.

\bibitem[AW11]{AW11}
Eric Allender and Fengming Wang.
\newblock On the power of algebraic branching programs of width two.
\newblock In {\em ICALP (1)}, pages 736--747, 2011.

\bibitem[Bar89]{bar}
D.A. Barrington.
\newblock Bounded-width polynomial size branching programs recognize exactly
  those languages in {NC$^1$}.
\newblock {\em Journal of Computer and System Sciences}, 38:150--164, 1989.

\bibitem[BCS97]{bcs-book97}
P.~B{\"u}rgisser, M.~Clausen, and M.A. Shokrollahi.
\newblock {\em Algebraic Complexity Theory}.
\newblock Springer, 1997.

\bibitem[BK09]{BK09}
Ir{\'e}n{\'e}e Briquel and Pascal Koiran.
\newblock A dichotomy theorem for polynomial evaluation.
\newblock In {\em MFCS}, pages 187--198, 2009.

\bibitem[BKM11]{BKM11}
Ir{\'e}n{\'e}e Briquel, Pascal Koiran, and Klaus Meer.
\newblock On the expressive power of cnf formulas of bounded tree- and
  clique-width.
\newblock {\em Discrete Applied Mathematics}, 159(1):1--14, 2011.

\bibitem[Bl{\"a}13]{Blaser13}
Markus Bl{\"a}ser.
\newblock Noncommutativity makes determinants hard.
\newblock In {\em Proceedings of ICALP}, volume 7965 of {\em Lecture Notes in
  Computer Science}, pages 172--183. Springer, 2013.
\newblock ECCC TR 2012-142.

\bibitem[BOC92]{benor.cleve}
M.~Ben-Or and R.~Cleve.
\newblock Computing algebraic formulas using a constant number of registers.
\newblock {\em SIAM Journal on Computing}, 21:54--58, 1992.

\bibitem[Bod98]{Bodlaender-tcs98}
Hans~L. Bodlaender.
\newblock A partial k-arboretum of graphs with bounded treewidth.
\newblock {\em Theoretical Computer Science}, 209(1–2):1 -- 45, 1998.

\bibitem[Bre74]{Brent}
R~P Brent.
\newblock The parallel evaluation of general arithmetic expressions.
\newblock {\em Journal of the ACM}, 21:201--206, 1974.

\bibitem[B{\"u}r00a]{burgisser-book00}
P.~B{\"u}rgisser.
\newblock {\em Completeness and Reduction in Algebraic Complexity Theory},
  volume~7 of {\em Algorithms and Computation in Mathematics}.
\newblock Springer, 2000.

\bibitem[B{\"u}r00b]{burgisser-tcs00}
Peter B{\"u}rgisser.
\newblock Cook's versus {V}aliant's hypothesis.
\newblock {\em Theor. Comput. Sci.}, 235(1):71--88, 2000.

\bibitem[B{\"u}r09]{Bur-cc09}
Peter B{\"u}rgisser.
\newblock On defining integers and proving arithmetic circuit lower bounds.
\newblock {\em Computational Complexity}, 18(1):81--103, 2009.

\bibitem[Cai90]{Cai90}
Jin{-}yi Cai.
\newblock A note on the determinant and permanent problem.
\newblock {\em Inf. Comput.}, 84(1):119--127, 1990.

\bibitem[CCL10]{CCL10}
Jin{-}yi Cai, Xi~Chen, and Dong Li.
\newblock Quadratic lower bound for permanent vs. determinant in any
  characteristic.
\newblock {\em Computational Complexity}, 19(1):37--56, 2010.

\bibitem[CDM13]{CDM13}
Florent Capelli, Arnaud Durand, and Stefan Mengel.
\newblock The arithmetic complexity of tensor contractions.
\newblock In {\em STACS}, volume~20 of {\em LIPIcs}, pages 365--376, 2013.

\bibitem[Che04]{Cheng04}
Qi~Cheng.
\newblock On the ultimate complexity of factorials.
\newblock {\em Theor. Comput. Sci.}, 326(1-3):419--429, 2004.

\bibitem[Csa76]{csa}
L.~Csanky.
\newblock Fast parallel inversion algorithm.
\newblock {\em SIAM J of Computing}, 5:818--823, 1976.

\bibitem[Dam91]{dam91}
C.~Damm.
\newblock \mbox{DET=L$^{({\rm \#L})}$}.
\newblock Technical Report Informatik--Preprint 8, Fachbereich Informatik der
  Humboldt--Universit\"at zu Berlin, 1991.

\bibitem[DM11]{DM11}
Arnaud Durand and Stefan Mengel.
\newblock On polynomials defined by acyclic conjunctive queries and weighted
  counting problems.
\newblock {\em CoRR}, abs/1110.4201, 2011.

\bibitem[DMPY12]{DMPY12}
Zeev Dvir, Guillaume Malod, Sylvain Perifel, and Amir Yehudayoff.
\newblock Separating multilinear branching programs and formulas.
\newblock In {\em STOC}, pages 615--624, 2012.

\bibitem[dMS96]{ms-ams96}
W.~de~Melo and B.F. Svaiter.
\newblock The cost of computing integers.
\newblock {\em Proceedings of the American Mathematical Society},
  124(5):1377--1378, 1996.

\bibitem[FLMS13]{FLMS13}
Herv{\'e} Fournier, Nutan Limaye, Guillaume Malod, and Srikanth Srinivasan.
\newblock Lower bounds for depth 4 formulas computing iterated matrix
  multiplication.
\newblock {\em Electronic Colloquium on Computational Complexity (ECCC)},
  20:100, 2013.

\bibitem[FMM12]{FMM12}
Herv{\'e} Fournier, Guillaume Malod, and Stefan Mengel.
\newblock Monomials in arithmetic circuits: Complete problems in the counting
  hierarchy.
\newblock In {\em STACS}, pages 362--373, 2012.

\bibitem[GKKP11]{GKKP11}
Bruno Grenet, Erich Kaltofen, Pascal Koiran, and Natacha Portier.
\newblock Symmetric determinantal representation of weakly-skew circuits.
\newblock In {\em STACS}, pages 543--554, 2011.

\bibitem[GKKS13a]{GKKS-ccc13}
Ankit Gupta, Pritish Kamath, Neeraj Kayal, and Ramprasad Saptharishi.
\newblock Approaching the chasm at depth four.
\newblock In {\em IEEE Conference on Computational Complexity}, 2013.

\bibitem[GKKS13b]{GKKS-focs13}
Ankit Gupta, Pritish Kamath, Neeraj Kayal, and Ramprasad Saptharishi.
\newblock Arithmetic circuits: A chasm at depth three.
\newblock In {\em Foundations of Computer Science (FOCS), IEEE}, 2013.
\newblock ECCC 2013-026.

\bibitem[GMT13]{GMT13}
Bruno Grenet, Thierry Monteil, and St{\'e}phan Thomass{\'e}.
\newblock Symmetric determinantal representations in characteristic 2.
\newblock {\em Linear Alg. Appl.}, 439(5):1364--1381, 2013.

\bibitem[Gre12a]{Grenet-thesis}
Bruno Grenet.
\newblock {\em Repr{\'e}sentation des polyn{\^o}mes, algorithmes et bornes
  inférieures}.
\newblock PhD thesis, \'Ecole Normale Supérieure de Lyon, 2012.

\bibitem[Gre12b]{Grenet12}
Bruno Grenet.
\newblock An upper bound for the permanent versus determinant problem.
\newblock manuscript, 2012.

\bibitem[Hya79]{Hyafil79}
Laurent Hyafil.
\newblock On the parallel evaluation of multivariate polynomials.
\newblock {\em SIAM J. Comput.}, 8(2):120--123, 1979.

\bibitem[Jan08]{Jan08}
Maurice~J. Jansen.
\newblock Lower bounds for syntactically multilinear algebraic branching
  programs.
\newblock In {\em MFCS}, volume 5162 of {\em Lecture Notes in Computer
  Science}, pages 407--418. Springer, 2008.

\bibitem[JMR12]{JMR12}
M~Jansen, M~Mahajan, and B~V~Raghavendra Rao.
\newblock Resource trade-offs in syntactic multilinear arithmetic circuits.
\newblock {\em Computational Complexity}, page to appear, 2012.
\newblock preliminary versions in MFCS 2008 and CSR 2009.

\bibitem[Kal85]{Kalorkoti85}
K.~Kalorkoti.
\newblock A lower bound for the formula size of rational functions.
\newblock {\em SIAM J. Comput.}, 14(3):678--687, 1985.

\bibitem[Kay10]{kayal-eccc10}
Neeraj Kayal.
\newblock Algorithms for arithmetic circuits.
\newblock {\em Electronic Colloquium on Computational Complexity (ECCC)},
  17:73, 2010.

\bibitem[KK08]{KK08}
Erich Kaltofen and Pascal Koiran.
\newblock Expressing a fraction of two determinants as a determinant.
\newblock In {\em ISSAC}, pages 141--146. ACM, 2008.

\bibitem[Koi05]{Koiran05}
Pascal Koiran.
\newblock Valiant's model and the cost of computing integers.
\newblock {\em Computational Complexity}, 13(3-4):131--146, 2005.

\bibitem[Koi10]{Koiran-dagstuhl10}
Pascal Koiran.
\newblock Complexity of arithmetic circuits (a skewed perspective).
\newblock In {\em Slides from Dagstuhl seminar 10481}. DROPS, 2010.
\newblock
  http://www.dagstuhl.de/Materials/Files/10/10481/10481.KoiranPascal.Slides.pdf.

\bibitem[Koi12]{Koiran12}
Pascal Koiran.
\newblock Arithmetic circuits: The chasm at depth four gets wider.
\newblock {\em Theor. Comput. Sci.}, 448:56--65, 2012.

\bibitem[Mal03]{Malod-thesis}
G.~Malod.
\newblock {\em Polynoˆmes et coefficients}.
\newblock PhD thesis, Univ. Claude Bernard – Lyon 1, 2003.

\bibitem[Men11]{mengel11}
Stefan Mengel.
\newblock Characterizing arithmetic circuit classes by constraint satisfaction
  problems - (extended abstract).
\newblock In {\em ICALP (1)}, pages 700--711, 2011.

\bibitem[Mes89]{meshulam}
Roy Meshulam.
\newblock On two extremal matrix problems.
\newblock {\em Linear Algebra and its Applications}, 114–115:261 -- 271,
  1989.
\newblock <ce:title>Special Issue Dedicated to Alan J. Hoffman</ce:title>.

\bibitem[MM61]{mm61}
M.~Marcus and H.~Minc.
\newblock On the relation between the determinant and the permanent.
\newblock {\em Illinois Journal of Mathematics}, 5:376--381, 1961.

\bibitem[Mor97]{moreira-ams97}
Carlos Gustavo T. de~A. Moreira.
\newblock On asymptotic estimates for arithmetic cost functions.
\newblock {\em Proceedings of the American Mathematical Society}, 125(2):pp.
  347--353, 1997.

\bibitem[MP08]{MP08}
Guillaume Malod and Natacha Portier.
\newblock Characterizing valiant's algebraic complexity classes.
\newblock {\em J. Complexity}, 24(1):16--38, 2008.

\bibitem[MR04]{MR04}
Thierry Mignon and Nicolas Ressayre.
\newblock A quadratic bound for the determinant and permanent problem.
\newblock In {\em International Mathematics Research Notices}, pages
  2004--4241, 2004.

\bibitem[MV97]{mahvin}
M.~Mahajan and V~Vinay.
\newblock Determinant: combinatorics, algorithms, complexity.
\newblock {\em Chicago Journal of Theoretical Computer Science {\tt
  http://www.cs.uchicago.edu/publications/cjtcs}}, 1997:5, dec 1997.
\newblock Preliminary version in Proceedings of the {\em Eighth Annual ACM-SIAM
  Symposium on Discrete Algorithms} SODA 1997, 730--738.

\bibitem[P\'13]{polya1913}
G.~P\'olya.
\newblock Aufgabe 424.
\newblock {\em Archiv der Mathematik und Physik (3)}, 20:271, 1913.

\bibitem[Rao10]{rao-thesis}
B~V~Raghavendra Rao.
\newblock {\em A Study of Width Bounded Arithmetic Circuits and the Complexity
  of Matroid Isomorphism}.
\newblock PhD thesis, The Institute of Mathematical Sciences, Chennai, India.,
  2010.
\newblock http://www.imsc.res.in/xmlui/handle/123456789/177.

\bibitem[Raz06]{raz-mlin-nc}
Ran Raz.
\newblock Separation of multilinear circuit and formula size.
\newblock {\em Theory of Computing}, 2(1):121--135, 2006.
\newblock preliminary version in FOCS 2004.

\bibitem[Raz09]{raz-det-perm}
Ran Raz.
\newblock Multi-linear formulas for permanent and determinant are of
  super-polynomial size.
\newblock {\em J. ACM}, 56(2), 2009.
\newblock preliminary version in STOC 2004.

\bibitem[Raz10]{raz-elusive}
Ran Raz.
\newblock Elusive functions and lower bounds for arithmetic circuits.
\newblock {\em Theory of Computing}, 6(1):135--177, 2010.

\bibitem[RSY08]{rsy08}
Ran Raz, Amir Shpilka, and Amir Yehudayoff.
\newblock A lower bound for the size of syntactically multilinear arithmetic
  circuits.
\newblock {\em SIAM J. Comput.}, 38(4):1624--1647, 2008.

\bibitem[RY08]{RY08-bal}
Ran Raz and Amir Yehudayoff.
\newblock Balancing syntactically multilinear arithmetic circuits.
\newblock {\em Computational Complexity}, 17(4):515--535, 2008.

\bibitem[RY09]{ry-cc09}
Ran Raz and Amir Yehudayoff.
\newblock Lower bounds and separations for constant depth multilinear circuits.
\newblock {\em Computational Complexity}, 18(2):171--207, 2009.

\bibitem[Rys63]{ryser1963}
H.J. Ryser.
\newblock {\em Combinatorial mathematics}.
\newblock Carus mathematical monographs. Mathematical Association of America,
  1963.

\bibitem[Sav70]{savitch70}
Walter~J. Savitch.
\newblock Relationships between nondeterministic and deterministic tape
  complexities.
\newblock {\em J. Comput. Syst. Sci.}, 4(2):177--192, 1970.

\bibitem[Str73]{strassen}
V.~Strassen.
\newblock Vermeidung von divisionen.
\newblock {\em Journal of Reine U. Angew Math}, 264:182--202, 1973.

\bibitem[SY10]{SY10-survey}
Amir Shpilka and Amir Yehudayoff.
\newblock Arithmetic circuits: A survey of recent results and open questions.
\newblock {\em Foundations and Trends in Theoretical Computer Science},
  5(3-4):207--388, 2010.

\bibitem[Tav13]{Tavenas13}
S{\'e}bastien Tavenas.
\newblock Improved bounds for reduction to depth 4 and depth 3.
\newblock In {\em MFCS}, volume 8087 of {\em Lecture Notes in Computer
  Science}, pages 813--824. Springer, 2013.

\bibitem[Tod92]{toda.circ}
S.~Toda.
\newblock Classes of arithmetic circuits capturing the complexity of computing
  the determinant.
\newblock {\em IEICE Transactions on Information and Systems}, E75-D:116--124,
  1992.

\bibitem[Val79]{valiant-stoc79a}
Leslie~G. Valiant.
\newblock Completeness classes in algebra.
\newblock In {\em STOC}, pages 249--261, 1979.

\bibitem[Val82]{valiant82}
L.~G. Valiant.
\newblock Reducibility by algebraic projections.
\newblock In {\em Logic and Algorithmic: International Symposium in honour of
  Ernst Specker}, volume~30, pages 365--380. Monograph. de l'Enseign. Math.,
  1982.

\bibitem[Val92]{val92}
L.~G. Valiant.
\newblock Why is boolean complexity theory difficult?
\newblock In M.~S. Paterson, editor, {\em Boolean Function Complexity}.
  Cambridge University Press, 1992.
\newblock London Mathematical Society Lecture Notes Series 169.

\bibitem[Ven92]{ve92}
H.~Venkateswaran.
\newblock Circuit definitions of nondeterministic complexity classes.
\newblock {\em SIAM J. on Computing}, 21:655--670, 1992.

\bibitem[Vin91]{vin91}
V~Vinay.
\newblock Counting auxiliary pushdown automata and semi-unbounded arithmetic
  circuits.
\newblock In {\em Proceedings of 6th Structure in Complexity Theory
  Conference}, pages 270--284, 1991.

\bibitem[Vol99]{vollmertext}
H.~Vollmer.
\newblock {\em Introduction to Circuit Complexity: A Uniform Approach}.
\newblock Springer-Verlag New York Inc., 1999.

\bibitem[VSBR83]{VSBR83}
Leslie~G. Valiant, Sven Skyum, S.~Berkowitz, and Charles Rackoff.
\newblock Fast parallel computation of polynomials using few processors.
\newblock {\em SIAM J. Comput.}, 12(4):641--644, 1983.

\bibitem[vzG87]{vzG87}
Joachim von~zur Gathen.
\newblock Permanent and determinant.
\newblock {\em Linear Algebra and its Applications}, 96(0):87 -- 100, 1987.

\end{thebibliography}
\end{document}